# Precission disease networks (PDN)

Javier Cabrera, Dhammika Amaratunga,
William Kostis and John Kostis

**Abstract**— This paper presents a method for building patient-based networks that we call Precision disease networks, and its uses for predicting medical outcomes.
Our methodology consists of building networks, one for each patient or case, that describes the disease evolution of the patient (PDN) and store the networks as a set of features in a dataset of PDN's, one per observation. We cluster the PDN data and study the within and between cluster variability. In addition, we develop data visualization technics in order to display, compare and summarize the network data. Finally, we analyze a dataset of heart diseases patients from a New Jersey statewide database MIDAS (Myocardial Infarction Data Acquisition System, in order to show that the network data improve on the prediction of important patient outcomes such as death or cardiovascular death, when compared with the standard statistical analysis.

## 1 INTRODUCTION

In modern medicine, networks have been used to represent pathways of diseases. In functional genomics researchers build biological pathways representing the mechanisms of certain processes. In cardiology and more generally in medicine, and in many other areas, researchers build graphs or pathways that summarize the working mechanisms of a certain process or disease.

In this paper we use the word network to represent a data construction in a form of a connected graph, representing the pathway of a disease. Note that in general this can model the pathway of any process. Such a network can then be analyzed to reveal medical insights. For example, in Neurology it is possible to build physical networks that represent real connections in a subject's brain and the dataset of brain networks have been analyzed using a Bayesian model [1] [2][3].

In cardiovascular diseases, while we do not have a physical network of connections, we do observe relationships. In this paper, we propose to build individual patient networks, which represent the disease evolution of patients across subsequent hospitalizations. These networks, like the neurology networks, are individual based and represent the evolutionary steps of various cardiovascular conditions, diseases and procedures.

Bayesian networks (BN) have been proposed in the past as tools for building networks. But the methodology of BN is to build one network for one dataset or subset that has been selected. This approach does not work here because each patient has a different disease progression and physicians believe that most cardiovascular diseases occur in many different ways. Therefore one single network or a top down approach may not capture this variability in most cases. Instead we propose the idea of building one network for each patient, which we term as personalized disease networks (PDN). As a result we obtain a dataset of PDN's. Then we cluster the PDN's by projecting to the principal components space and use the clusters as follows:

(i) Clusters are summarized as networks representing a diversity of disease evolution pathways for the groups of patients in our dataset.

(ii) Predict clinical outcomes such as 1 year or 5 years cardiovascular death, stroke, and other. We expect that different networks correspond to different survival probabilities or distributions.

- *J. Cabrera is the corresponding author and is with the Dept of Statistics and Biostatistics, Rutgers, The State University of New Jersey,110 FrelinghuysenRd, Piscataway, NJ08854. E-mail: cabrera@stat.rutgers.edu.*
- *D. Amaratunga, is with the Rutgers Robert Wood Johnson Medical School Cardiovascular Institute, Rutgers The State University of New Jersey, Suite 5200, 125 Paterson Str, New Brunswick, NJ08904. E-mail: damaratung@yahoo.com.*
- *W.Kostis, is with the Rutgers Robert Wood Johnson Medical School Cardiovascular Institute, Rutgers The State University of New Jersey, Suite 5200, 125 Paterson Str, New Brunswick, NJ08904. E-mail: wkostis@rwjms.rutgers.edu.*
- *J.Kostis is with the Rutgers Robert Wood Johnson Medical School Cardiovascular Institute, Rutgers The State University of New Jersey, Suite 5200, 125 Paterson Str, New Brunswick, NJ08904. E-mail: kostis@rwjms.rutgers.edu.*

## 2 THE MIDAS DATABASE AND AN EXAMPLE

The MIDAS database [4] contains all hospitalizations for cardiovascular diseases in New Jersey since. The MIDAS database contains the records of 15 million hospitalizations for 5 million patients with cardiovascular diseases. We apply our network methodology to MIDAS data and build patient networks for cardiovascular diseases. MIDAS contains the demographic information of the patients as well as the ICD9 codes for the diseases and the procedural codes.

Example: For this study we select a subset of MIDAS containing those patients who were admitted for an event of heart failure (HF) or atrial fibrillation (AFIB) or both. The medical experts selected a total of 36 conditions that maybe related to HF or AFIB. For simplicity we restricted ourselves to the subset of African American patients.

The first step of our methodology is to build the PDN's from the data. In our study, we use African American patient subset to precede the analysis.

## 3. BUILDING PDN'S.

Medical records in MIDAS include for each patient a list of all the hospitalizations together with dates and ICD9 codes that are diagnosed. These codes represent all the cardiovascular events and conditions that happened during the hospitalization. This information was used to build the edges of the PDN's following a set of rules.

Our network is made up of nodes 1,...,*p* that may be connected by directed edges from one node to another, which we will call arrows.Suppose that a patient has an episode of atrial fibrillation (AFIB) followed by an episode of heart failure (HF). If the time elapsed between both episodes is less than a threshold $\theta$, then we will add an arrow from the AFIB node to the HF node. If the reverse were to happen at a different point in time the opposite arrow will also be added to the network. The following algorithm implements this idea.

(i) Start with a data matrix $A$ containing n rows (patients) and $p$ columns (events) giving the dates of the events corresponding to each patient. If the i-th patient never had the j-th event the corresponding entry in the dataset should be empty or NA (not available).

(ii) For the i-th patient build a network (PDN) with nodes N1,...,N*k(i)*, representing the list of events that occurred to the i-th patient of length *k(i)* out of *p* and have a non empty entry in the i-th row the matrix A. If 0 < A[i,*l*]–A[i,j] < $\theta_{jl}$ then one edge of the network will connect nodes *j* and *l* with an arrow from node *j* to node *l*, otherwise there will be no connection. In this way, the adjacency network matrix *M* is built. $\theta_{jl}$ is a threshold that is provided by the medical experts or a combination of medical expertise with data driven methodldogy that will be discused later.

(iii) For every pair of nodes *j* and *l* step (ii) will be repeated in both directions so it is possible to have two arrows one from *j* to *l* and another from *l* to *j*. Also it is possible to have only one arrow or no arrows. The arrows between *j* and *l* are recorded by two dummy variables *j*to*l* and *l*to*j*, which will take values of 1 or 0 representing arrow or no arrow.

(iv) After repeating steps (ii) and (iii) for all *n* subjects we obtain a network matrix *M* of *n* rows and *p*×(*p*-1) columns, two columns for every pair of *I* and *J*. The rows of *M* represent the individual networks of the *n* subjects.

Fig. 1 depicts the result of the above algorithm applied to the data from patient no. 28 in our example dataset that is described in section 2. The list of events and dates for this patient is shown in the middle of the figure. The events are ordered by date of occurrence and the arrows represent the relationships satisfying a given set of rules.

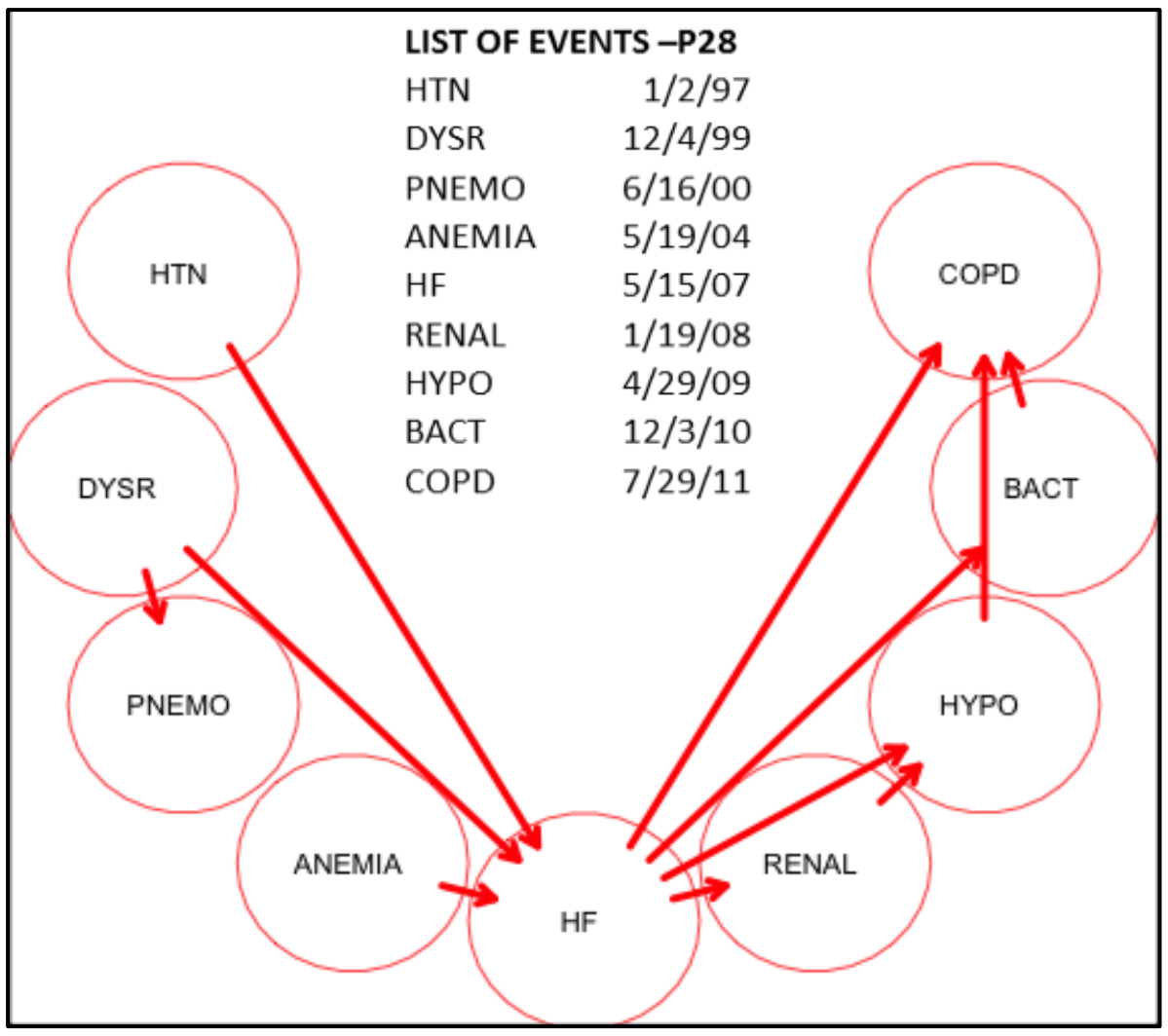

Fig.1.a Graph of PDN for patient number 28 on our dataset. The table in the middle of the graph shows the sequence of codes and dates for the paptient

## 4. CLUSTERING PDN'S AND DIMENSION REDUCTION.

The matrix *M* contains a diversity of networks that represent different ways of disease evolution or different pathways, one per subject. However, it may be possible to summarize the disease evolution using just a few network summaries that are followed by a large majority of the patients. One simple way to do this is to apply a hierarchical clustering or any equivalent algorithm to the subject's data, or rows of *M*. Since the dimensionality of *M* can be quite large it is recommended to reduce the dimension by using principal components analysis (PCA) prior to clustering [5]. In particular we apply enriched PCA which is a weighted version of PCA that gives less importance to variables whose relationship to the outcome variable is explained by chance [6]. If a response variable is available the weights s

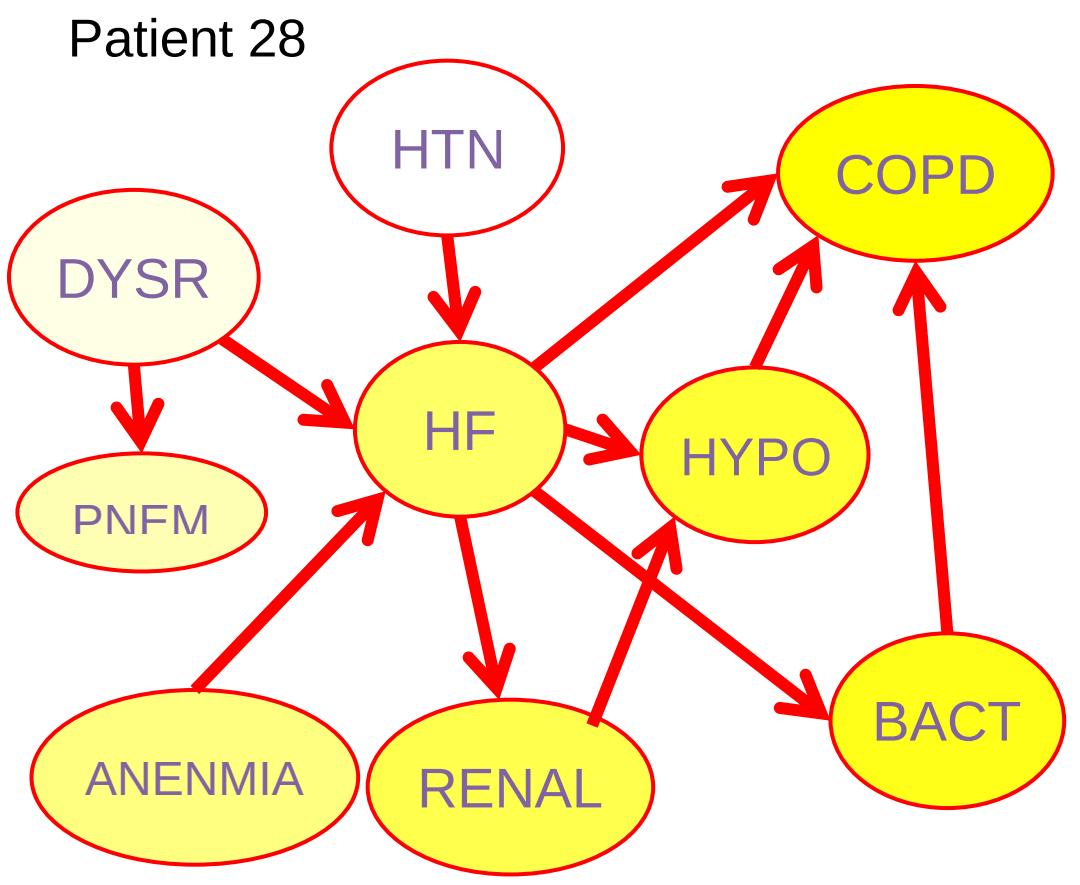

Fig 1.b has a idealization of the graph in Fig 1.a

## 5. Data analysis and Results

Our example dataset contains a total of 4500 patients that complied with our inclusion criteria. In MIDAS there are other subsets of interest with millions of subjects but for the purpose of illustrating the PDN methodology we consider only African American subjects that have been diagnosed with hear failure (HF), atrium fibrillation (AFIB) or both.

The objective of this example is to understand the relationship between AFIB and HF and the pathways in which AFIB precedes HF and the reverse in which HF precedes AFIB. For this example we consider 36 comorbidities including myocardial infarction (MI), stroke (STR), hypertension (HTN) and others. The picture in Figure 1 depicts the PDN for patient no. 28 in our datatset of 4500 PDN's. For this dataset we implemented the following data analysis plan:

1. Build the matrix of personalized disease networks according to the hospitalization records and the set of rules or thresholds that are obtained provided by the expert physicians or that are derived from the data.
2. Apply enriched PCA [6] or a nonlinear PCA method [7] to determine the principal components.
3. Perform the hierarchical clustering on a reduced set of principal components and screen for the optimal numbers of clusters.

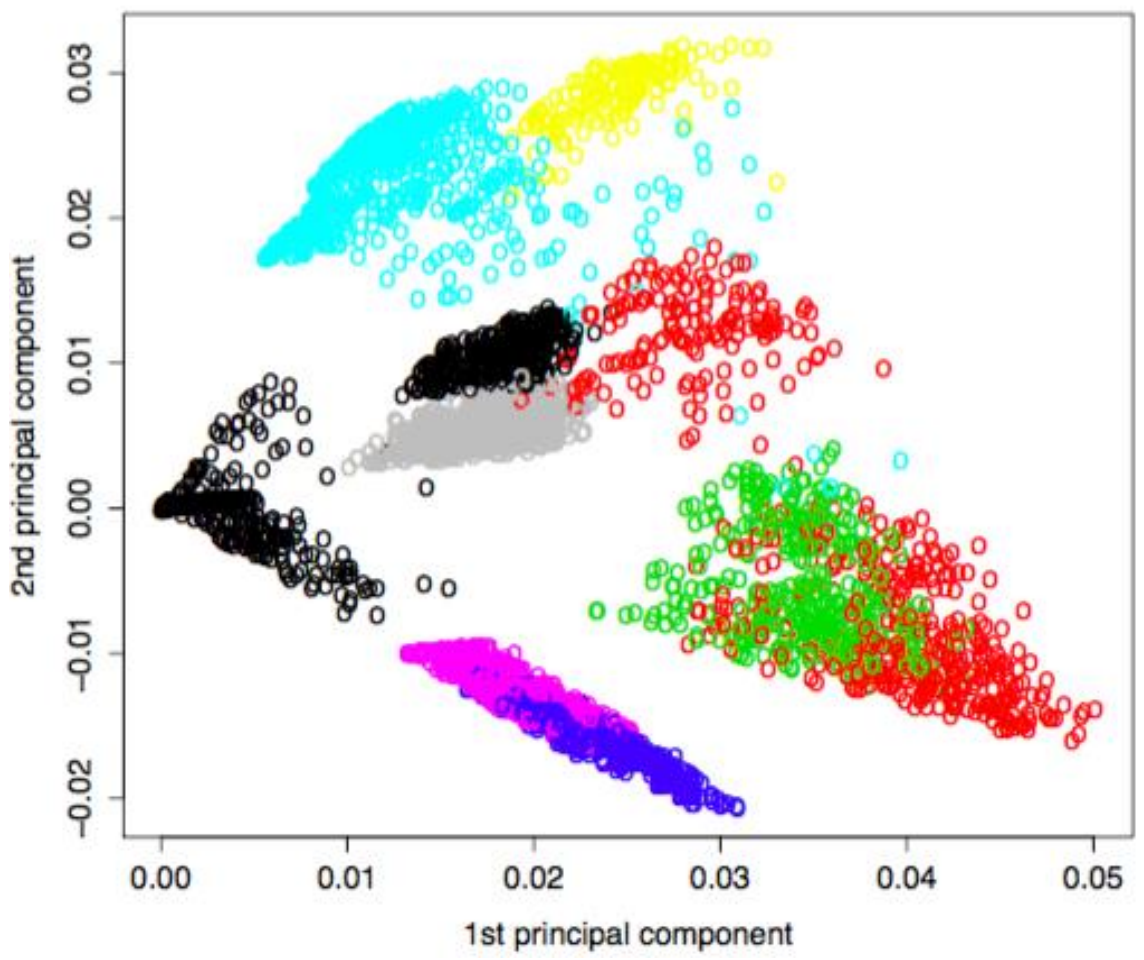

Fig.2 10 principal component clusters. PDNs were clustered using 43 enrichedprincipal components of the adjacent matrix. The clusters were plotted against the 1st and 2nd principal components.

Fit prediction models that could be linear, logistic, poisson or cox regression models depending on the type of respose. The predictors are grouped in 4 blocks which are (i) Age and Gender, (2) 36 comorbidities (3) $k$ PCA's summary of the set of PDN's and, (iv) Cluster factor. Table 1 shows the $R_2$ for 7 models of interest that are ciombinations of these groups. The two columns representing the results derived form the summary of the PDN dataset using Linear and non-linear PCA respectively.

5. Consider a hierarchy of models that combine datasets (i) through (iv) as the 7 models in Table 1 and perform likelihood ratio chi-square tests comparing models in the hierarchy in order to establish the importance of the PDN and cluster datasets ((iii) and (iv)).

In our study we obtained two versions of the PDN matrix $M_1$ using data driven thresholds, and $M_2$ using no threshold, that is setting all $\theta_{IJ}$'s to infinity. We performed parallel analyses for these two datasets.

For step 2 we considered two methods for PCA, enriched PCA [5] and nonlinear PCA[7]and from the SCREE plots we reduced the dimension to 43 principal componets for each of the two analyses of the 2 datasets which resulted in 4 pricipal componets datsets.

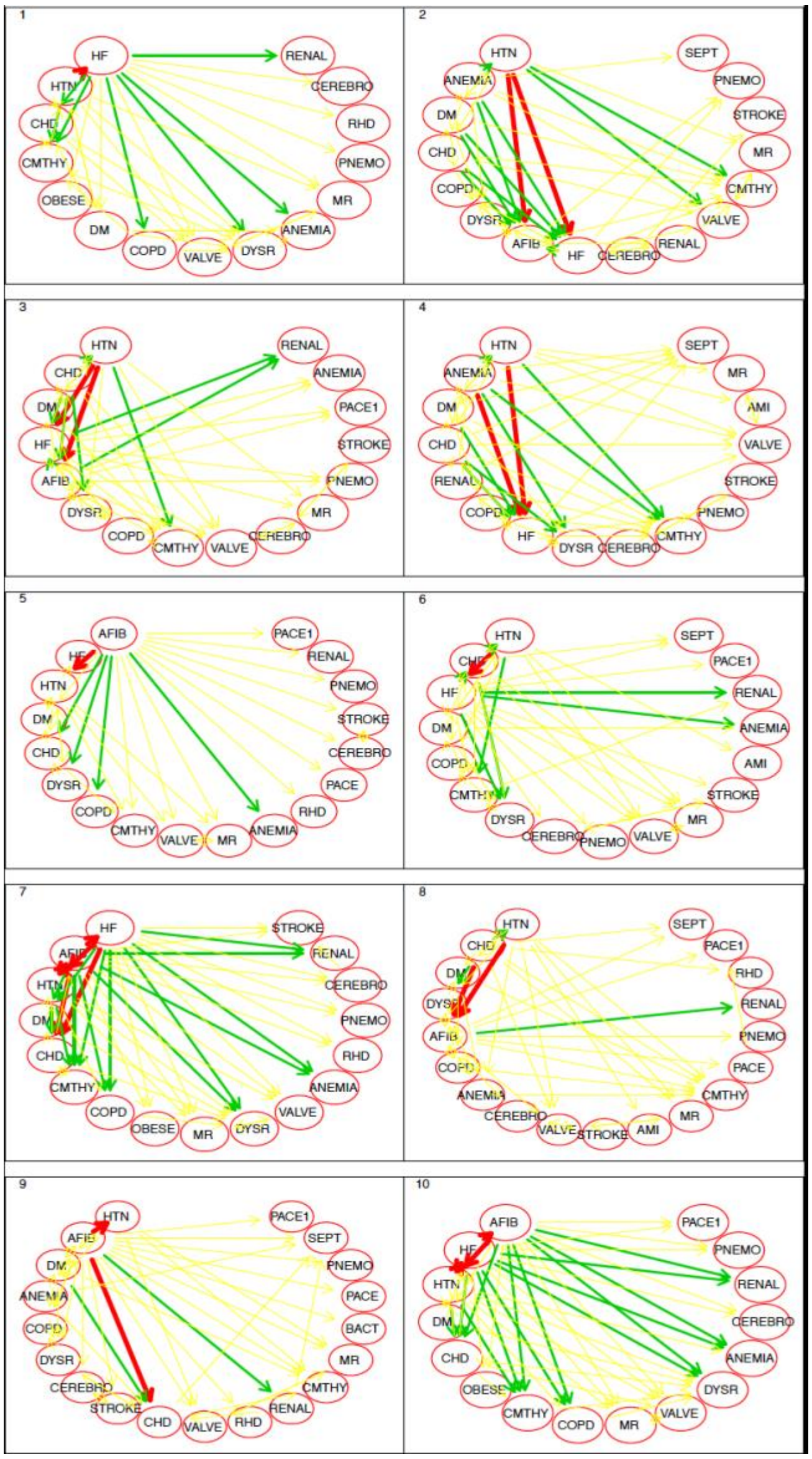

Fig.3. Summary graph of ten clusters of PDN's. The arrows represent the frequency of the relationship from A to B in the cluster of patients, Red arrows represente a frequency of 75% or more of the cluster observations containinig the relationship, green arrows represent a frequency in the range 50%- 75%, whereas yellow arrows represent represent a frequency in the range 25%-50%.

For step 3 we performed a hierarchical cluster analysis using the WARD method that resulted in 10 clusters for each of the 4 PCA datasets. Figure 2 shows a scatter plot of the first two components of the first of the cluster analysis where the 10 clusters are shown in different colors. In Figure 3 we display the summaries of the 10 clusters.

For step 4 we used the cox proportional hazard model with the variable "all death" as response that measures the date of death for any cause of death. We fitted 7 different models using different combinations of predictors as shown on table 1. The best $R^2$ (55%) was found for model 7 and dataset 2 that corresponds to enriched principal components on the network dataset constructed using the data driven thresholds.

Finally for the 5th step in table 1 we first compare models 4 and 1 and conclude that the addition of the cardiovascular events improves the fit from the model with just demographics. Then we show that the addition of the PDN variables also improves the fit not only in $R^2$ but also is statistically very significant. On table 1, the line of Model 6 vs model 5 shows that the addition of the cluster factor only improves the fit for the Nonlinear PCA datasets. Finally the results of model 7 vs model 5 shows that the use of clusters as a multiplicative variable results in statistically significant improvements for all but the linear PCA with infinity threshold.

In summary the best model is model 7 for the PCA dataset with data driven thresholding because it greatly improves the likelihood as compared to any other model and has a much higher $R^2$ than any of the other models.

## 7 Conclusion

The results shown here demonstrate the methodology of building personalized disease networks. They also show that PDN's improve the $R^2$ and goodness of fit with respect to the standard models and also show that the improvements are substancial in terms of $R^2$. Therefore, this establishes the effectiveness of PDNs, albeit in a preliminary study. We plan to explore this in the future.

## Acknowledgment

This research was supported by a Chancellor seed grant from Rutgers University-Newark.

### Table 1

Comparisons of 7 models to predict All Death in 4 different datasets of principal components derived from PDN datsets.

| | PCA no cut (11 clusters) | PCA with cut (10 clusters) | NLPCA no cut (10 clusters) | NLPCA with cut (10 clusters) |
|---|---|---|---|---|
| **R square** | | | | |
| **1. age+ sex** | 0.064 | 0.064 | 0.064 | 0.064 |
| **2. disease condition** | 0.11 | 0.11 | 0.11 | 0.11 |
| **3. network variables** | 0.27 | 0.25 | 0.27 | 0.25 |
| **4. age+ sex+ disease** | 0.15 | 0.15 | 0.15 | 0.16 |
| **5. age+sex+ disease+ network** | 0.28 | 0.28 | 0.28 | 0.28 |
| **6. age+sex+ disease+ network+ clusters** | 0.28 | 0.28 | 0.27 | 0.27 |
| **7. age+sex+ disease+ network * clusters** | 0.37 | 0.55 | 0.31 | 0.32 |
| **p-value** | | | | |
| **model 4/model 1** | 0 | 0 | 0 | 0 |
| **model 4/model 5** | 0 | 0 | 0 | 0 |
| **model 6/model 5** | 0.1048 | 0.6919 | 8.59E-13 | 6.59E-13 |
| **model 7/model 5** | 0.997 | 0 | 0.002567 | 0.000395 |